\documentclass{article}

\PassOptionsToPackage{numbers, sort, compress}{natbib}

\usepackage[dblblindworkshop, final]{neurips_2026}
\usepackage{amsmath}
\usepackage{enumitem}

\usepackage[utf8]{inputenc} 
\usepackage[T1]{fontenc}    
\usepackage{hyperref}       
\usepackage{url}            
\usepackage{booktabs}       
\usepackage{amsfonts}       
\usepackage{nicefrac}       
\usepackage{microtype}      
\usepackage{xcolor}         
\usepackage{xspace}
\usepackage[pdftex]{graphicx}
\graphicspath{{./}{figures/}}

\title{From Historical Schedules to On-Sky Control: Autonomous Scheduling on the 4-m Blanco Telescope}

\workshoptitle{9th Workshop on Machine Learning for the Physical Sciences (ML4PS), 2026}

\author{%
  Rachel Hur\\
  Department of Physics\\ University of Chicago\\ 5720 S.\ Ellis Ave., Chicago, IL 60637 \\
  NSF-Simons AI Institute for the Sky (SkAI)\\ 172 E. Chestnut St., Chicago, IL 60611 \\ 
  \texttt{rhur@uchicago.edu} \\
  \And
  Paul Chichura \\
  NSF-Simons AI Institute for the Sky (SkAI)\\ 172 E. Chestnut St., Chicago, IL 60611 \\
  Department of Astronomy \& Astrophysics\\ University of Chicago\\ 5640 S.\ Ellis Ave., Chicago, IL 60637 \\
  \texttt{pchichura@uchicago.edu} \\
  \And
  Alex Drlica-Wagner \\
  Fermi National Accelerator Laboratory\\ P.O.\ Box 500, Batavia, IL 60510\\
  Department of Astronomy \& Astrophysics\\ University of Chicago\\ 5640 S.\ Ellis Ave., Chicago, IL 60637 \\
  NSF-Simons AI Institute for the Sky (SkAI)\\ 172 E. Chestnut St., Chicago, IL 60611 \\
  \texttt{kadrlica@fnal.gov} \\
  \And
  Eric~H.~Neilsen~Jr. \\
  Fermi National Accelerator Laboratory\\ P.O.\ Box 500, Batavia, IL 60510\\
  \texttt{neilsen@fnal.gov} \\
  \And
  Guillermo Damke  \\
  Cerro Tololo Inter-American Observatory/NSF NOIRLab, Casilla 603, La Serena, Chile\\
  \texttt{guillermo.damke@noirlab.edu} \\
  \And
  Aleksandra Ciprijanovic \\
  Fermi National Accelerator Laboratory\\ P.O.\ Box 500, Batavia, IL 60510, USA \\
  Department of Astronomy \& Astrophysics\\ University of Chicago\\ 5640 S.\ Ellis Ave., Chicago, IL 60637 \\
  NSF-Simons AI Institute for the Sky (SkAI)\\ 172 E. Chestnut St., Chicago, IL 60611 \\
  \texttt{aleksand@fnal.gov} 
}

\begin{document}

\begin{flushright}
    FERMILAB-CONF-26-0695-CSAID
\end{flushright}

\maketitle
\begin{abstract}
    The scientific efficiency of ground-based astronomical observatories relies on effective decision-making in a dynamic, multi-timescale environment with complex scientific rewards. Traditional observation schedulers are designed and tuned by domain experts to handle these considerations using heuristic rules based on models of the environment and observational system. Here, we leverage historical data to train an automated telescope scheduler using behavior cloning (BC). The BC scheduler is trained purely on historical datasets; it is not given expert scheduling heuristics nor an optimal scientific reward. It is then evaluated in order of increasing realism: first on expert states, then in closed-loop offline rollouts, and finally during autonomous operation with real-time feedback on the NSF's Victor M.\ Blanco 4-m Telescope. On expert states from holdout nights, the BC policy closely matches the expert distribution over airmass and Moon separation. During closed-loop rollouts on the same nights, we find that the per-observation data-quality metric, $t_{\rm eff}$, falls by $6.9 \pm 3.4\%$ relative to the expert's. During on-sky deployment, the policy scheduled $149$ exposures in $5.2$ hours with an average $t_{\rm eff} = 0.75 \pm 0.26$, consistent with closed-loop rollout evaluations. No additional degradation was observed when transitioning to real observing conditions. This represents the first real-world, real-time deployment of a learned autonomous scheduler on a leadership-class national observatory and establishes a path toward on-sky evaluation of reinforcement-learning-based schedulers trained on archival schedules augmented with counterfactual trajectories.
\end{abstract}

\section{Introduction}

Astronomical survey scheduling, i.e., deciding what a telescope observes and when, is a high-dimensional and multi-objective sequential decision-making problem within a partially stochastic environment. The problem is akin to the well-known traveling salesman problem, a combinatorial optimization problem that is known to be NP-hard \citep{korte2018combinatorial}.
Astronomical survey scheduling is further complicated by a changing environment (weather, sky brightness, target availability), high-priority interrupts, and long-term survey goals, which jointly impact the value assigned to each decision. Furthermore, since each decision changes the subsequent set of feasible observations, locally favorable decisions do not necessarily produce globally optimal solutions \citep{2019PASP..131k5002R}.
Traditional telescope schedulers \citep{neilsen2019_obstac, 2016SPIE.9910E..13D, 2019AJ....157..168D, 2021ApJS..256....2D} address these challenges using iteratively tuned heuristic rules and priorities that are based on models of the environment and observing system. They generally incorporate human expertise, physical models, and explicit weighting of short- and long-term scientific objectives.
Recent astronomical surveys provide extensive historical data capturing scheduling decisions and their consequences under real observing conditions, offering a foundation for learned schedulers that reduce reliance on computationally expensive simulations and human expertise.

Machine-learning (ML) approaches to astronomical scheduling have increasingly explored the use of simulated training environments, including reinforcement learning (RL) for resource allocation and cooperative scheduling across telescope arrays \cite{zhang2024}, target-of-opportunity and follow-up scheduling \cite{wang25, zhang25}, and autonomous target selection \cite{terranova23, cao2025, zhang26}. For the Legacy Survey of Space and Time (LSST) \cite{lsst}, Naghib et al. \cite{naghib19} explored RL and global optimization strategies to learn optimal weights for a set of expert-designed basis functions.
While these studies demonstrate the promise of ML-driven scheduling, they do not jointly address the large action spaces, long survey horizons, real-time observing feedback, and iterative on-sky deployment considered here.

In this work, we train a model to {\it directly learn} scheduling behavior from archival scheduling data generated and executed by an expert-designed algorithm, \texttt{obstac} \cite{neilsen2019_obstac}. These data comprise six years of Dark Energy Survey (DES; \citep[][]{2016MNRAS.460.1270D}) observations obtained with the Dark Energy Camera (DECam; \citep[][]{flaugher2015_decam}) on the NSF's Victor M.\ Blanco 4-m Telescope at Cerro Tololo Inter-American Observatory in Chile.
This methodological approach anchors our algorithm in real observing conditions and reduces dependence on a fully synthetic and computationally expensive training environment, but also restricts coverage to trajectories only present in historical data.
We use behavior cloning (BC) \citep{bain99} to assess what imitation alone recovers, where it breaks down, and whether its performance differs when transitioning to on-sky deployment. We evaluate the scheduler in three increasingly realistic regimes: (1) single-step evaluation on expert states held out during training, (2) closed-loop policy rollouts on held-out test nights where each policy decision determines the next simulated environment state, and (3) real-time scheduling on the Blanco telescope. For this, we develop a modular, policy-independent framework that connects learned schedulers to real-time telemetry and the Blanco telescope control systems.

\vspace{-0.1cm}
\section{Data and Methods}

{\bf Training and Evaluation Data.}
The dataset is compiled from six years of archival DES wide-field observations, which cover ${\sim} 5{,}000 \deg^2$ of the southern sky over 758 nights using ${\sim} 1{,}700$ target fields. Each field was visited ${\sim}10$ times in each of five broadband filters (\textit{grizY}) using a pre-defined dithering strategy where the target pointing for each field is offset \citep{diehl_desy6_2020, 2021ApJS..255...20A}. Operational metadata was assembled from the mountain-top telemetry database \citep{2008SPIE.7019E..11H}. Nights that contain only time-domain observations (which utilize a scheduling strategy distinct from wide-field observations) are removed from the dataset, leaving 689 nights. Because state--action pairs within each night are strongly correlated, the data are split randomly at the night level using a train/validate/test split of $80\%/10\%/10\%$ corresponding to $553/68/68$ nights consisting of $63{,}824/8{,}417/7{,}978$ state--action pairs.\footnote{Training and evaluation code is available at \url{https://github.com/rachelhur/skyway}, and analysis and plotting scripts are available at \cite{zenodo_2026_24401148}.}

\vspace{0.1em}{\bf State and Action Space.}
Each action $a \in A$ ($|A|=30{,}400$) is composed of a $(pointing, filter)$ pair, where $pointing$ is the sky coordinates in the topocentric horizontal frame (i.e., Az/El) and $filter$ is one of the five DES broadband filters ($grizY$).
The pointing is discretized into $N_{\rm side}=32$ HEALPix pixels \cite{2005ApJ...622..759G},\footnote{\url{http://healpix.sourceforge.net}} which we call ``bins,'' and we restrict the action space to the 6,080 bins that are above the horizon. By using topocentric horizontal coordinates rather than equatorial coordinates, the policy is capable of generalizing to targets outside the DES footprint.
The state vector $s \in S$ contains current environmental features: the telescope pointing and filter, the Sun and Moon position, per-band sky brightness predicted with \texttt{skybright} \cite{skybright}, the seeing predicted from the measured PSF FWHM of exposures taken in the preceding 15 minutes, and the per-band survey progress. The state also includes per-bin features that describe each candidate action: airmass, slew distance from current pointing, Sun and Moon distance, time until bin sets, and local survey-completion fraction.

\vspace{0.1em}{\bf Modeling the Policy.}
\label{ssec:policy}
The policy $\pi_\theta(a|s)$ is a conditional probability distribution over actions. We parameterize the policy using a shared-weight multi-layer perceptron (MLP). The global state is encoded into a 128-dimensional representation and concatenated with each candidate bin's features before being passed through the shared-weight, three-layer MLP with 128-dimensional hidden layers, LeakyReLU activations, and layer normalization. The network outputs a score per candidate action, and the scores of invalid candidate actions are masked before the score vector is sent through the softmax layer. The highest probability action (i.e., pointing--filter pair), $a^*=(b^*,f^*)$, is selected. Invalid pointing--filter combinations that do not contain an unfinished field are masked, resulting in a state-dependent action space, $A(s) \subseteq A$. After an action is selected, if there are multiple target fields in the selected action, only fields that require the filter $f^*$ associated with the chosen action are retained, and we interpolate the network scores for that filter from the HEALPix bin centers to the candidate field positions using \texttt{Scipy}'s \texttt{RBFInterpolator} with a quintic kernel \cite{scipy2020}. The scheduler then selects the field with the highest score.

\vspace{0.1em}{\bf Training.}
The policy is trained using BC \cite{bain99}, a maximum-likelihood supervised learning technique. The archival data are used to fit MLP parameters $\theta$ of policy $\pi_\theta$. Because our action space is discrete, we minimize cross-entropy between the expert actions and the policy, $\pi_\theta(a|s)$, where $\pi_\theta$ is normalized over the set of viable actions $A(s) \subseteq A$. We train with \texttt{Adam} using a batch size of 1024, an initial learning rate of $10^{-3}$ with cosine annealing scheduler, and early stopping based on validation loss. The model checkpoint is selected from the validation split. Training takes ${\sim}30$~min on an H100 GPU \citep{delta_ai}. 

\vspace{-0.1cm}
\section{Results and Discussion}

As a data quality summary metric, we use the effective exposure time scale factor,\\[-1.2em]

\begin{equation}\label{eq:teff}
    t_{\rm eff} = \eta^2 \left(\frac{0.9''}{\rm FWHM} \right)^2 \left(\frac{b_{\rm dark}}{b} \right)
\end{equation}

where $\eta$ is atmospheric transmission (fraction of light that makes it to the telescope), $\rm FWHM$ is the full-width at half maximum of the point spread function (i.e., seeing), and $b$ and $b_{\rm dark}$ denote the observed and nominal reference dark sky brightness, respectively \cite{neilsen2016_teff, neilsen2019_obstac}. The $t_{\rm eff}$ of an exposure describes the achieved sensitivity compared to that expected in nominal observing conditions (e.g., a 10\% improvement in $t_{\rm eff}$ is equivalent to $\sim$10\% improvement in operational efficiency).

\vspace{0.1em}{\bf Expert State Evaluation.}
On single-step predictions from expert holdout states, the learned policy matches the expert's filter $61.4 \%$ of the time, approximately $3\times$ greater than uniform selection across five filters. Marginal per-filter distributions of sky brightness and Moon elevation (two key factors in the DES scheduler's algorithm determining filter choice) closely reproduce those of the expert (see App.~\ref{appx:ss} for details).
The learned policy reproduces the expert's pointing bin $24.3\%$ of the time, ${\sim}10^{3}\times$ more frequently than a uniform selection across bins, and lands within $5\deg$ of the expert's field in $81.3\%$ of transitions, with a median separation of $2.08\deg$. 
Averaged over the same transitions, the learned policy agrees with the expert airmass ($1.189$ against $1.190$), Moon separation ($92.7\deg$ against $92.8\deg$), $t_{\rm eff}$ ($0.672$ against $0.672$), and slews substantially less per exposure ($1.65\deg$ against $4.72\deg$).
The DES scheduler often chose $g$- and $r$-band filters interchangeably in similar observing conditions, as well as $z$- and $Y$-band filters. A similar degeneracy exists in pointing choice, where multiple bins in a neighborhood were deemed equally favorable. We therefore do not expect even a well-trained scheduler to reproduce DES actions exactly.

\begin{figure}[thb!]
    \centering
    \includegraphics[width=1.\linewidth]{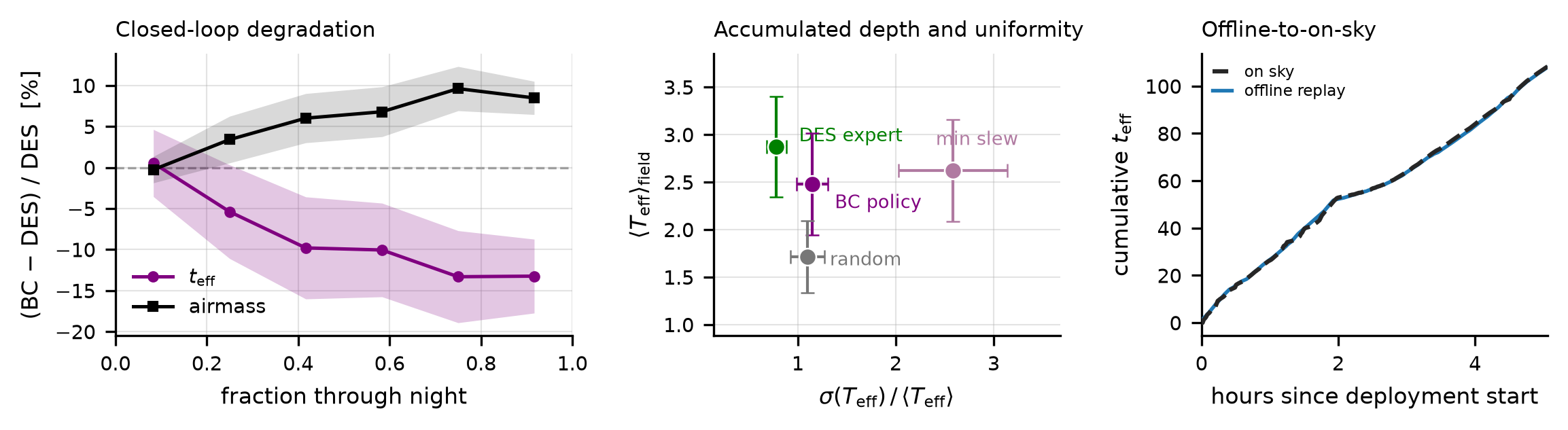}
    \caption{(Left) Percentage difference in mean $t_{\rm eff}$ (purple) and mean airmass (black) between the learned BC scheduler during closed-loop rollout and the DES scheduler on holdout nights. The mean is calculated per bin, per night before averaging over nights. Bands show the $95\%$ interval from a delete-one-night jackknife over test nights. 
    (Middle) Accumulated depth against uniformity for the DES scheduler, BC scheduler, and two heuristics (uniform random and minimum slew). $T_{\rm eff}$ is the $t_{\rm eff}$ summed over each field's exposures from all holdout nights. The y-axis is its mean over footprint fields and the x-axis is its standard deviation across fields divided by that mean, so smaller values mean more uniform depth across the footprint. Unvisited fields contribute $T_{\rm eff}=0$. Error bars are $95\%$ intervals from jackknife resampling over test nights. (Right) The cumulative $t_{\rm eff}$ over all previous observations as a function of hours since the start of deployment. The on-sky deployment (dashed, black) closely agrees with the estimated offline replay (solid, blue).}\label{fig:mp}
\end{figure}

\vspace{0.1em}{\bf Offline Closed-Loop Behavior.} The learned policy is rolled out over all 68 test nights, initialized on the first state of each night; one night with too few expert exposures is excluded from the comparisons below.
Because simulated schedules have no measured data quality, we estimate $t_{\rm eff}$ for both the expert and offline closed-loop rollout schedules from the conditions measured during that night (see App.~\ref{appx:teff} and \citep{neilsen2016_teff, neilsen2019_obstac} for details).
The two schedules diverge early in the night for all test nights and never resynchronize, so direct exposure-level comparison is not possible. 
We therefore compare by-night and estimate uncertainties from jackknife resampling.
On average, per-exposure $t_{\rm eff}$ of the learned scheduler falls by $6.9 \pm 3.4 \%$ relative to the expert.
The deficit grows with time through the night (Fig.~\ref{fig:mp}, left): the difference is consistent with zero at the start of the night and reaches $\sim 13\%$ towards the end of the night. It correlates with a drift toward setting targets and higher airmass that only appears in closed-loop rollouts. On expert states, the policy closely reproduces the expert's airmass and hour-angle distributions (App.~\ref{appx:ms}).
This pattern is suggestive of compounding error \citep{ross11}: the deployed policy's preference for short slews is already present on expert states and may keep the scheduler observing a region as it sets, carrying it to higher airmass than the expert observes on the same night (see App.~\ref{appx:ms} for further discussion).

The DES scheduler prioritized both accumulated $t_{\rm eff}$ as well as uniform coverage across the survey footprint over time. We compare the DES expert against closed-loop rollouts of the learned policy and non-learned baseline heuristics on test nights in Fig.~\ref{fig:mp} (middle panel). To measure both depth and uniformity, we sum $t_{\rm eff}$ over all test-night exposures of each field $i$ in the footprint, $T_{{\rm eff},i} = \sum t_{\rm eff}$, with unvisited fields set to $T_{{\rm eff},i}=0$. The vertical axis is the mean over all fields, $\langle T_{\rm eff}\rangle$, and the horizontal axis is the fractional spread $\sigma(T_{\rm eff})/\langle T_{\rm eff}\rangle$, where $\sigma(T_{\rm eff})$ is the standard deviation of $T_{{\rm eff},i}$ across fields. A lower fractional spread means more uniform sky coverage, so better schedules lie toward the upper left. The ``random'' heuristic randomly chooses observable fields, while the ``min slew'' heuristic simply chooses the closest unfinished field. These baselines represent extreme ends of the tradeoff between optimizing for $\langle T_{\rm eff}\rangle$ and covering the sky uniformly. The BC scheduler accumulates $\sim$50\% more $\langle T_{\rm eff}\rangle$ than random selection and covers the footprint much more uniformly than the min slew heuristic. The scheduler produces comparable accumulated $\langle T_{\rm eff}\rangle$ relative to the expert, but falls behind in its ability to spread out observations over the footprint.

\vspace{0.1em}{\bf On-Sky Deployment.}
The learned policy was used to control Blanco autonomously for the second half-night on 24 June 2026, running for ${\sim}5.2$~hours from 4:48 to 10:00 UTC. It executed $149$ exposures in $griz$, and produced an average $t_{\rm eff} = 0.75$ with a standard deviation of $0.26$. The scheduler had the option to perform long slews and slews to high airmass, but consistently chose reasonable targets with high effective exposure time.
To compare whether this differs from the expected offline evaluation, we replay the same night offline using the same checkpoint, candidate fields, and measured seeing trajectory. The two runs observe the same number of fields, and the offline replay observed $92\%$ of the field--filter pairs observed on-sky.
Both the offline replay and on-sky deployment deliver a mean per-exposure $t_{\rm eff}$ of $0.75$. For this single half-night of observing, we detect no additional degradation when moving from offline replay to on-sky deployment, indicating agreement between offline and on-sky performance.
This agreement is consistent with the interpretation that the closed-loop degradation on test nights reflects declining expert-cloning fidelity rather than simulator error.

\vspace{0.1em}{\bf Limitations.}
Each bin is approximately the size of the DECam field of view (${\sim}3$~deg$^2$), so our candidate actions do not carry information about neighboring fields, which is necessary for learning an optimal tiling policy.
Future work will incorporate this information using architectures such as graph neural networks or a multi-level binning scheme to encode summary statistics about the neighborhood.
The effective exposure time scale factor, $t_{\rm eff}$, is not a complete description of scientific utility of a scheduler; other factors such as uniformity of depth and filter balance are also important and will be further considered.
Finally, the autonomous on-sky deployment is only a single night with particularly good weather conditions for the season. Robustness across different nights, seasons, and weather conditions requires evaluation in these regimes with additional on-sky deployments. However, we find that for the particular deployment night, the simulator and $t_{\rm eff}$ estimator are reliable predictors of on-sky performance.

\vspace{-0.1cm}
\section{Conclusion and Future Work}

We present analyses of a scheduler trained on six years of DES observations with BC in progressively more realistic conditions. 
Single-step predictions on expert states closely reproduce the archived schedules' distributions over filter choice, airmass, Moon separation, and $t_{\rm eff}$, but they predict a smaller average slew distance between pointings. 
On closed-loop rollouts, the average $t_{\rm eff}$ falls $6.9\%$ below the expert's as the policy drifts to higher airmass.
Initial on-sky performance is consistent with offline expectations, and suggests minimal sim-to-real gap for the conditions during the 5.2-hour, good-weather observing period.

This real-data baseline is intended to support ongoing work: training a scheduler with RL using archival observing nights augmented with counterfactual trajectories, whose states and rewards can be estimated using analytical projections that have been validated against real data.
Learning from trajectories that stray from, but are close to, expert-derived trajectories avoids a brute-force exploration of the large action space, and RL should reduce the closed-loop drift seen here by training the policy on states produced by its own decisions.
Comparing these future policies against the BC baseline will allow us to evaluate how performance changes as counterfactual experience is introduced, both offline and on-sky.


\begin{ack}

    We gratefully acknowledge the support of the NSF-Simons AI Institute for the Sky (SkAI) via grants NSF AST-2421845 and Simons Foundation MPS-AI-00010513. This work was supported with funding by the University of Chicago AI Initiative.  This work was supported in part by the U.S.\ Department of Energy, Office of Science, Office of High Energy Physics as part of the Genesis Mission. 
    
    This research used the DeltaAI advanced computing and data resource, which is supported by the National Science Foundation (award OAC 2320345) and the State of Illinois, and the Delta advanced computing and data resource which is supported by the National Science Foundation (award OAC 2005572) and the State of Illinois. Delta and DeltaAI are joint efforts of the University of Illinois Urbana-Champaign and its National Center for Supercomputing Applications.
    
    This work was produced by Fermi Forward Discovery Group, LLC under Contract No. 89243024CSC000002 with the U.S. Department of Energy, Office of Science, Office of High Energy Physics. Publisher acknowledges the U.S. Government license to provide public access under the DOE Public Access Plan.
    (\url{http://energy.gov/downloads/doe-public-access-plan}).

    This project used public archival data from the Dark Energy Survey (DES). Funding for the DES Projects has been provided by the U.S. Department of Energy, the U.S. National Science Foundation, the Ministry of Science and Education of Spain, the Science and Technology Facilities Council of the United Kingdom, the Higher Education Funding Council for England, the National Center for Supercomputing Applications at the University of Illinois at Urbana–Champaign, the Kavli Institute of Cosmological Physics at the University of Chicago, the Center for Cosmology and Astro-Particle Physics at the Ohio State University, the Mitchell Institute for Fundamental Physics and Astronomy at Texas A\&M University, Financiadora de Estudos e Projetos, Fundação Carlos Chagas Filho de Amparo à Pesquisa do Estado do Rio de Janeiro, Conselho Nacional de Desenvolvimento Científico e Tecnológico and the Ministério da Ciência, Tecnologia e Inovação, the Deutsche Forschungsgemeinschaft and the Collaborating Institutions in the Dark Energy Survey.
    
    The Collaborating Institutions are Argonne National Laboratory, the University of California at Santa Cruz, the University of Cambridge, Centro de Investigaciones Energéticas, Medioambientales y Tecnológicas–Madrid, the University of Chicago, University College London, the DES-Brazil Consortium, the University of Edinburgh, the Eidgenössische Technische Hochschule (ETH) Zürich, Fermi National Accelerator Laboratory, the University of Illinois at Urbana-Champaign, the Institut de Ciències de l’Espai (IEEC/CSIC), the Institut de Física d’Altes Energies, Lawrence Berkeley National Laboratory, the Ludwig-Maximilians Universität München and the associated Excellence Cluster Universe, the University of Michigan, the National Optical Astronomy Observatory, the University of Nottingham, The Ohio State University, the OzDES Membership Consortium, the University of Pennsylvania, the University of Portsmouth, SLAC National Accelerator Laboratory, Stanford University, the University of Sussex, and Texas A\&M University.
    
    Based in part on observations at Cerro Tololo Inter-American Observatory, National Optical Astronomy Observatory, which is operated by the Association of Universities for Research in Astronomy (AURA) under a cooperative agreement with the National Science Foundation.

    Some of the results in this paper have been derived using the healpy and HEALPix package.
\end{ack}





\bibliography{main}{}
\bibliographystyle{plainnat}
\clearpage
\appendix
\section{Single-step expert state evaluations}\label{appx:ss}

\begin{figure}[htb!]
    \centering
    \includegraphics[width=0.53\linewidth]{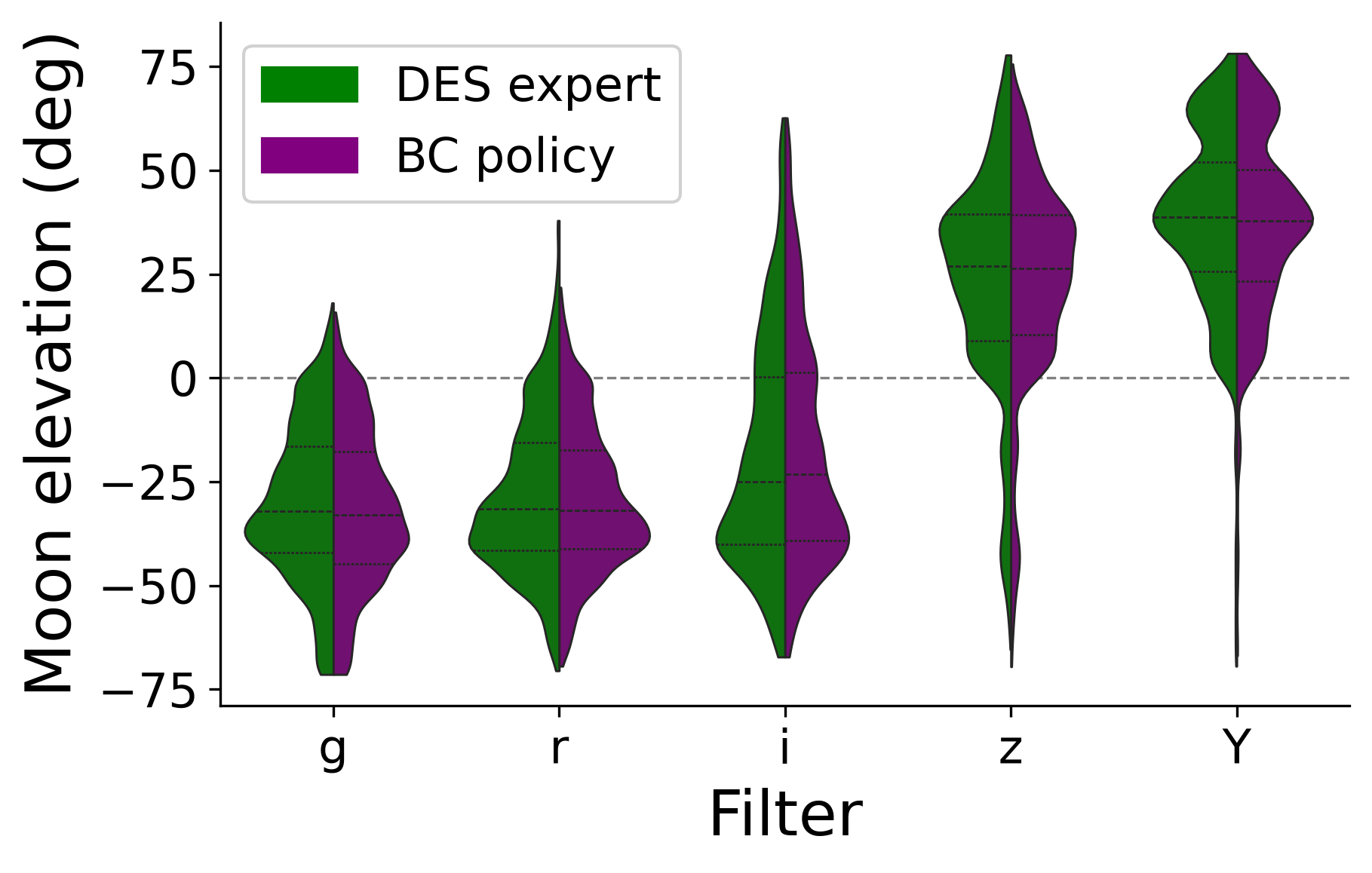}
    \includegraphics[width=0.40\linewidth]{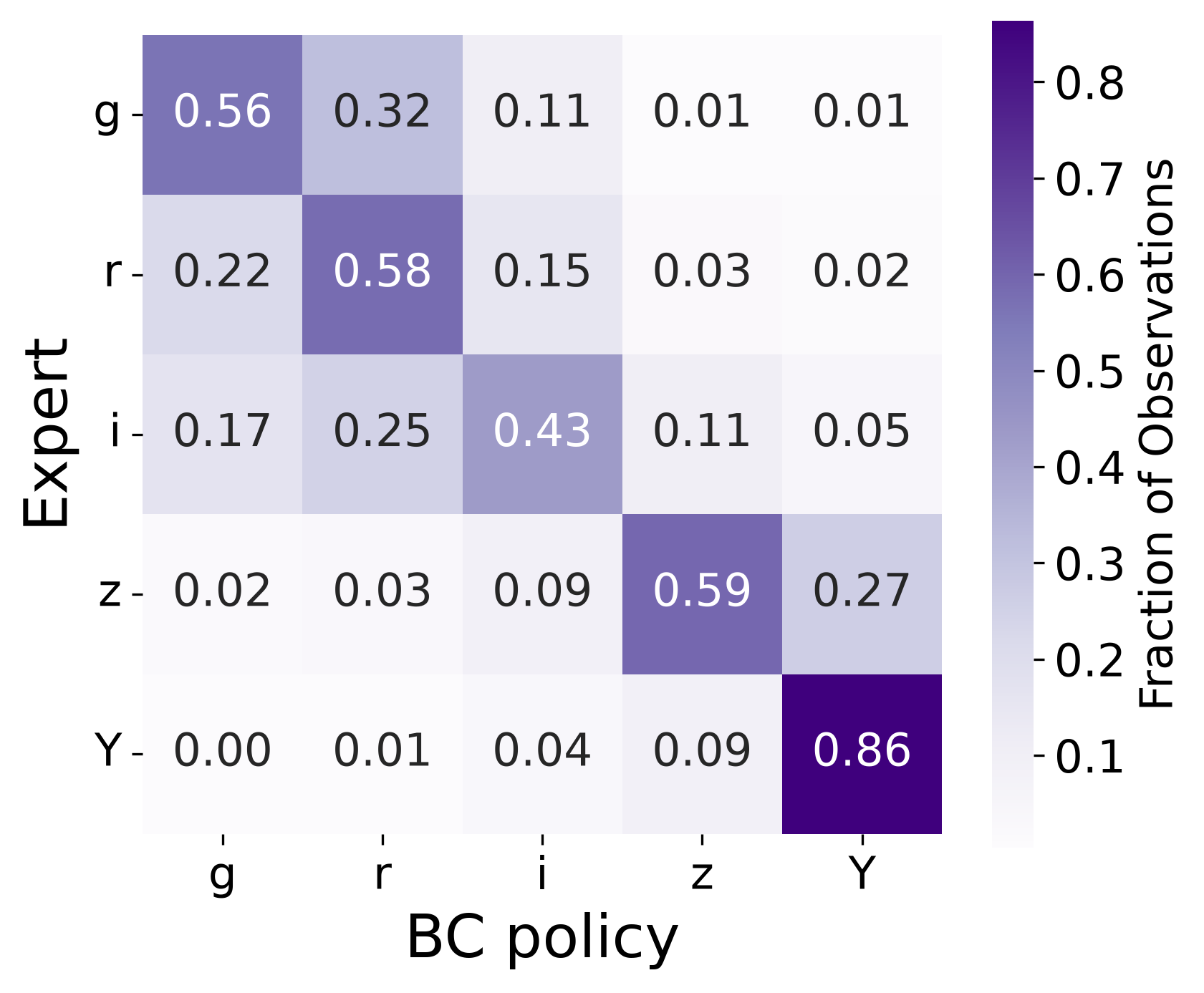}
    \caption{Filter choice on holdout expert states. (Left) Violin plots of the Moon elevation at each transition, split by selected filter, with expert DES scheduler in green and the BC policy in purple. (Right) BC filter choice against DES filter choice on the same states, normalized by row. Disagreements concentrate on the $g$/$r$ and $z$/$Y$ pairs, where the DES scheduler often chose either filter interchangeably.}
    \label{fig:violin}
\end{figure}

We evaluate filter choice fidelity using marginal distributions across Moon elevation, a strong factor that informed the DES scheduler's filter choice (Fig.~\ref{fig:violin}, left panel). Both schedulers show a similar distribution of Moon elevation, in both quantiles and overall shape of the distribution, for each filter. In the right panel of Fig.~\ref{fig:violin}, disagreements concentrate on the $g$/$r$ and $z$/$Y$ pairs, where the DES scheduler often chose either filter interchangeably.

\begin{figure}[htb!]
    \centering
    \includegraphics[width=1.\linewidth]{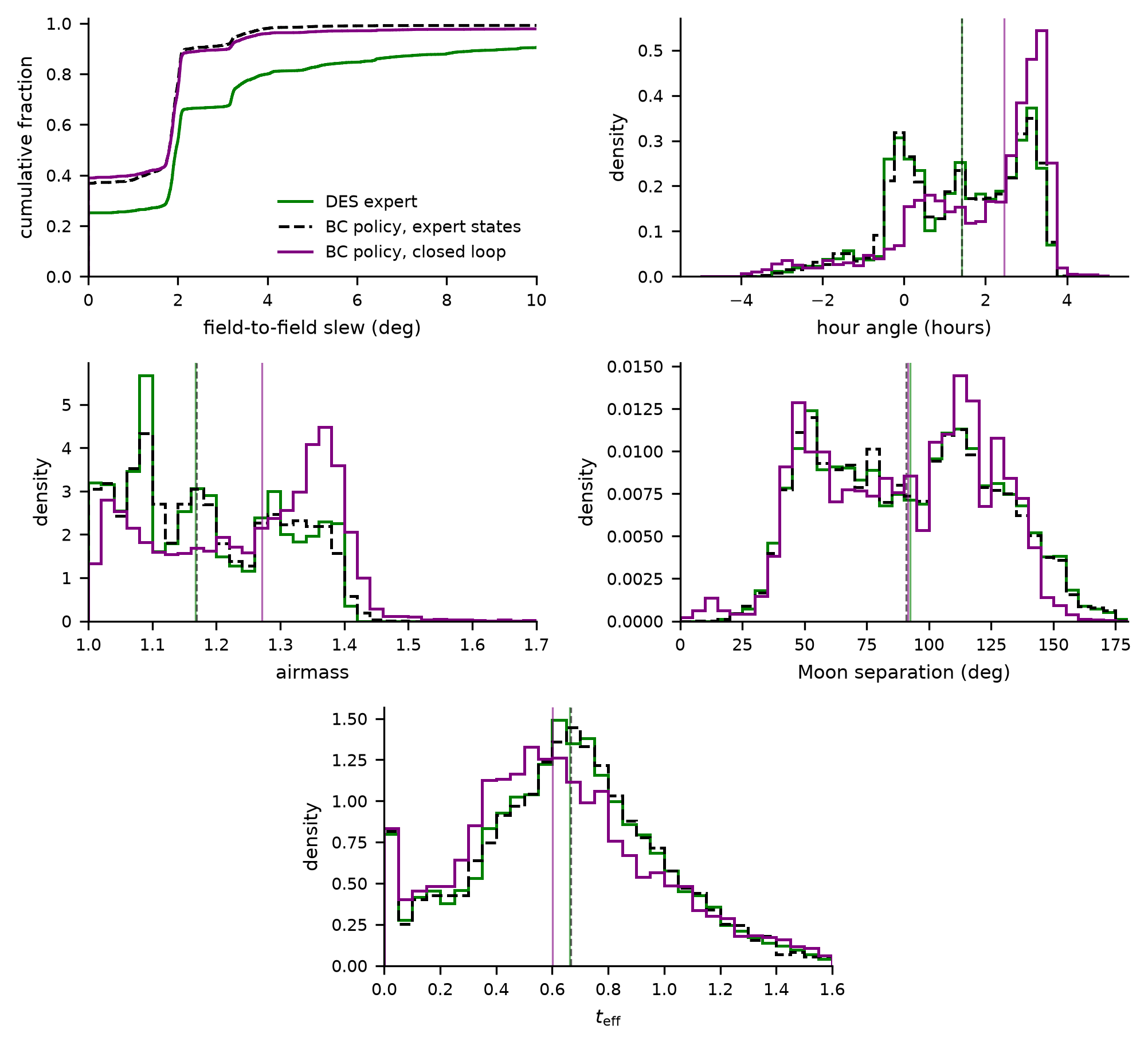}
    \caption{Pointing choices of the DES scheduler (green) and the BC scheduler on expert states (black, dashed) and in closed-loop rollout (purple, solid), on holdout nights, and the $t_{\rm eff}$ they deliver. (Top left) Cumulative distribution of field-to-field slew distance. (Top right) Hour angle of each exposure (negative is east, rising; positive is west, setting). (Middle left) Airmass. (Middle right) Moon separation. (Bottom) Estimated $t_{\rm eff}$ (App.~\ref{appx:teff}). Vertical lines mark each distribution's median.}
    \label{fig:pointing}
\end{figure}

Since sky pointing is spread across 6,080 bins and many bins have equal value, we do not rely on top-$k$ agreement, but rather marginal distributions over slew distance, hour angle, airmass, and Moon separation (Fig.~\ref{fig:pointing}, top and middle rows) and $t_{\rm eff}$ (Fig.~\ref{fig:pointing}, bottom) to evaluate pointing choice fidelity. The BC policy (black, dashed) and the DES expert (green) closely match for hour angle, airmass, Moon separation, and $t_{\rm eff}$. However, for slew distance, while both the BC policy and DES expert collect most of their exposures at slews less than ${\sim}2\deg$ and the overall shapes of the cumulative distributions are similar, the DES scheduler distributes exposures across a much larger range of slew.

\section{Closed-loop offline rollouts}\label{appx:ms}

{\bf Protocol}. Each of the 68 test nights is rolled out from the first expert state, i.e., from the first exposure taken by the DES scheduler. The historical data contain interruptions due to weather and other science programs that our environment simulator does not model. We identify these interrupts as gaps lasting $> 5$\,min (the maximum slew time for Blanco is ${\sim}4$~min for a 100\,deg slew). To avoid biasing the weather-dependent image quality comparison, we remove observations from the BC scheduler during these periods.

\begin{figure}[htb!]
    \centering
    \includegraphics[width=0.5\linewidth]{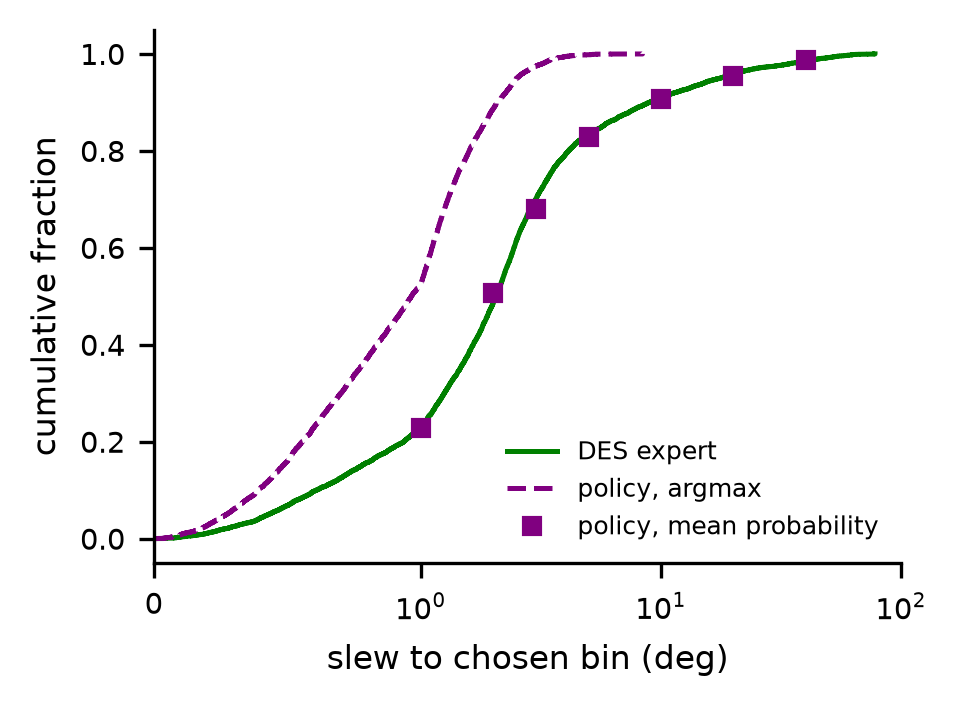}
    \caption{Cumulative distribution of slews on expert states, measured from the current pointing to the center of the chosen sky bin, for the DES expert (green) and the BC policy selecting its most probable action (argmax; purple, dashed). Purple squares show the cumulative slew distribution predicted by the BC policy's action probabilities, averaged over all expert states and evaluated at selected slew distances.}
    \label{fig:slew-calibration}
\end{figure}

{\bf Closed-loop drift}. We examine how the per-step bias present on expert states evolves during closed-loop rollout. The BC scheduler's slew distribution in closed-loop operation is nearly identical to its distribution when evaluated on single-step expert states, but both make far fewer long slews than the expert (Fig.~\ref{fig:pointing}, top left). Its airmass distribution, on the other hand, follows the expert's on single-step predictions but not in closed-loop rollout (Fig.~\ref{fig:pointing}, middle left).
The policy's airmass excess and $t_{\rm eff}$ deficit both start near zero and grow through the night, reaching ${\sim}9\%$ and ${\sim}13\%$ by the end of the night (Fig.~\ref{fig:mp}, left), and the closed-loop $t_{\rm eff}$ distribution shifts as a whole rather than only in its tail (Fig.~\ref{fig:pointing}, bottom).
On expert states the policy reproduces the expert's hour-angle distribution, but in closed loop it shifts toward setting targets (Fig.~\ref{fig:pointing}, top right).  
Together, these results are consistent with compounding error: by rarely making long slews, the policy may fail to reposition as its target region sets, which carries it to higher airmass than the expert observes on the same night.
Preliminary checks suggest this bias is amplified by always selecting the most probable action (Fig.~\ref{fig:slew-calibration}). On expert states, the slew distribution implied by the policy's own probabilities closely follows the expert's, whereas the slews it selects by taking the most probable action are much shorter. Although the policy assigns long slews a non-negligible share of probability, it is uncertain about the precise action, so a nearby bin is almost always the single most probable choice. Furthermore, this probability is spread over many distant bins, exacerbating the effect on longer slews.

{\bf Baseline construction}. The two non-learned heuristics are given only the night's observing window and the remaining number of visits required for each field.
At each decision step, they draw a random filter from \textit{grizY} and choose among fields that still require visits in that filter and lie below an airmass of $1.8$. The ``random'' heuristic draws a field assuming a uniform distribution across all fields, while ``min slew'' chooses the field closest to the previous pointing.
Each heuristic is evolved over the observing night using the same simulator as the closed-loop scheduler. 

\section{Estimating $\boldsymbol{t_{\rm eff}}$}\label{appx:teff}

The archival schedules record a measurement of each quantity that determines the effective exposure time scale factor (Eq.~\ref{eq:teff}) and $t_{\rm eff}$ itself.
In order to compare the performance of the learned scheduler directly against the expert on holdout nights, we reconstruct $t_{\rm eff}$ for both the archival schedules and learned scheduler using the archived data quality measurements, a set of analytical expressions \cite{neilsen2016_teff, neilsen2019_obstac}, and a model fit to the sky brightness excess \cite{neilsen2019_obstac}. For each component of $t_{\rm eff}$ (seeing, sky brightness, and cloud attenuation), the night's measured conditions are interpolated in time from all DECam exposures that night, and then evaluated at each scheduled exposure's time, pointing, and filter. Archival exposures are treated identically but using leave-one-out measurements.

\textbf{Sky brightness} is reconstructed from the $\texttt{skybright}$ model, $\Delta \mu_\texttt{skybright}$ plus a measured-minus-modeled residual of that night's exposures interpolated in time, $\Delta \mu_r$, which absorbs the slowly drifting offset between the model and DECam measurements. The residual is interpolated per filter from same-filter exposures within an hour when available, and otherwise from all filters with a fitted per-filter offset\\[-1.2em]

\begin{equation}\label{eq:sky}
    \Delta \mu(t, \alpha, \delta, f) = \Delta \mu_\texttt{skybright}(t, \alpha, \delta, f) + \Delta \mu_r(t, f)
\end{equation}

where $t$ is time, $\alpha, \delta$ are the Right Ascension and Declination sky coordinates, and $f$ is the filter. The sky brightness factor can then be calculated as\\[-1.2em]

\begin{equation}
    \frac{b_{\rm dark}}{b} = 10^{-0.4\,\Delta \mu}
\end{equation}

\textbf{Seeing} reconstruction follows the analytical expressions described in \cite{neilsen2019_obstac}. First, the instrumental term $0.45''$ is removed from each exposure's measured FWHM in quadrature and is projected to an $i$-band, zenith-equivalent value (directly overhead the observer). For each scheduled exposure, the interpolated value is re-projected to the scheduled band and airmass, and recombined in quadrature with the instrumental term. The airmass scaling is $(X_f/X_i)^{3/5}$ and the band-scaling is $(\lambda_f/\lambda_i)^{-0.34}$, where we fit the exponent to DECam band ratios. We use a fiducial FWHM of $0.9''(\lambda_f/\lambda_i)^{-1/5}$.

\textbf{Atmospheric transmission} follows DES conventions \cite{morganson2018}. The attenuation is modeled as \\[-1.2em]

\begin{equation}
    \Delta m = k_f (X - 1) + c(t)
\end{equation}

where $k_f$ is the per-band extinction coefficient, $X$ is the airmass, and $c(t)$ is the night's interpolated cloud attenuation. Attenuation below a 0.2 mag threshold is treated as clear sky \\[-1.2em]

\begin{equation}\label{eq:eta}
    \eta^2 =
    \begin{cases}
        1, & \Delta m \le 0.2, \\
        10^{-0.8\,(\Delta m - 0.2)}, & \Delta m > 0.2.
    \end{cases}
\end{equation}

Because cloud attenuation is only measured at the archival pointing, we assume that clouds are uniform across the sky.

\begin{figure}
    \centering
    \includegraphics[width=1.\linewidth]{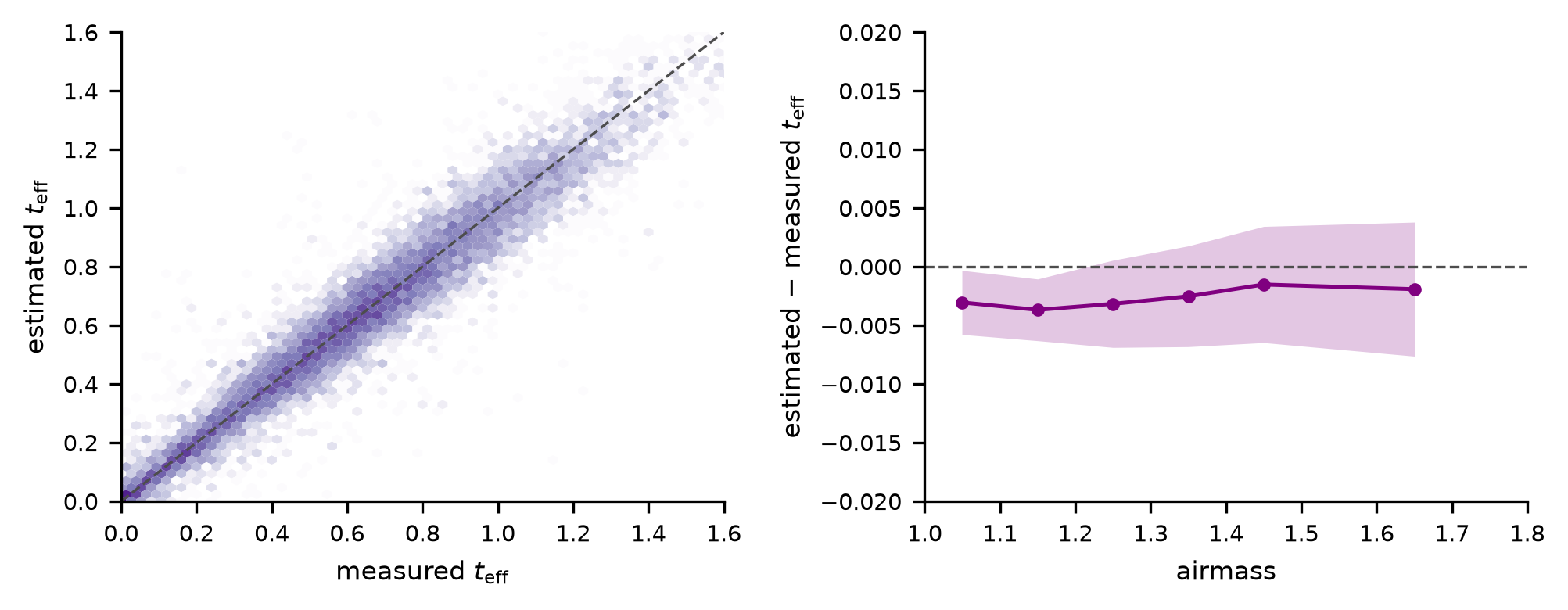}
    \caption{(Left) Logarithmic density of estimated vs.\ measured $t_{\rm eff}$, with the one-to-one line (black, dashed). (Right) Mean residual of estimated-minus-measured $t_{\rm eff}$ in airmass bins; the band shows the $95\%$ interval from jackknife resampling.}
    \label{fig:teff}
\end{figure}

\textbf{Validation}. We compare the leave-one-out estimate with the measured $t_{\rm eff}$ on test nights (Fig.~\ref{fig:teff}, left) and find that the estimate has a bias of $\Delta t_{\rm eff} = -0.003$, which is small compared with its NMAD scatter of $0.064$. The bias does not depend on airmass. Because clouds are assumed to be uniform across the sky, a schedule that points into locally cloudier (clearer) regions would not be penalized (rewarded).

\end{document}